\documentclass[prl,twocolumn,nofootinbib,showkeys,showpacs,superscriptaddress]{revtex4-1}
\usepackage{amssymb}
\usepackage{amsmath}
\usepackage{amstext}
\usepackage{amsfonts}
\usepackage{bbold}
\usepackage{bm}
\usepackage{graphics}
\usepackage{mathtools}
\usepackage{lmodern}
\usepackage{graphicx}
\usepackage[capitalise]{cleveref}

\newcommand{\be}{\begin{equation}}
\newcommand{\ee}{\end{equation}}
\newcommand{\la}{\langle}
\newcommand{\ra}{\rangle}
\newcommand{\sint}{{\textstyle\int}} 

\begin{document}

\title{The gravitational mass carried by sound waves} 

\author{Angelo~Esposito}
\affiliation{Department of Physics, Center for Theoretical Physics, Columbia University, 538W 120th Street, New York, NY, 10027, USA}
\affiliation{INFN, Sezione di Roma, Piazzale A. Moro 2, I-00185 Rome, Italy}
\affiliation{Theoretical Particle Physics Laboratory (LPTP), Institute of Physics, EPFL, Lausanne, Switzerland}

\author{Rafael~Krichevsky}
\affiliation{Department of Physics, Center for Theoretical Physics, Columbia University, 538W 120th Street, New York, NY, 10027, USA}

\author{Alberto~Nicolis}
\affiliation{Department of Physics, Center for Theoretical Physics, Columbia University, 538W 120th Street, New York, NY, 10027, USA}

\begin{abstract}
We show that the commonly accepted statement that sound waves do not transport mass is only true at linear order.  Using effective field theory techniques, we confirm the result found in \cite{Nicolis:2017eqo} for zero-temperature superfluids, and extend it to the case of solids and ordinary fluids. We show that, in fact, sound waves do carry mass---in particular, gravitational mass. This implies that a sound wave not only is affected by gravity but also generates a tiny gravitational field, an aspect not appreciated thus far.
Our findings are valid for non-relativistic media as well, and could have intriguing experimental implications.
\end{abstract}


\maketitle


\noindent\emph{\textbf{Introduction}} --- It is usually said that sound waves do not transport mass. They carry momentum and energy, 
but it is an accepted fact that the net mass transported by a sound wave vanishes.
Here, we question this ``fact".

A first indication that sound waves can carry a nonzero net mass is contained in the results of \cite{Nicolis:2017eqo}: there, using an effective point-particle theory, it was shown that phonons in zero-temperature superfluids have an effective coupling to gravity, which depends solely on their energy and on the superfluid's equation of state. For ordinary equations of state, this coupling corresponds to a {\em negative} effective gravitational mass: in the presence of an external gravitational field, such as that of Earth, a phonon's trajectory bends upwards. 

Now, this effect is completely equivalent to standard refraction: in the presence of gravity, the pressure of the superfluid depends on depth, and so does the speed of sound. As a result, in the geometric acoustics limit sound waves do not propagate along straight lines. Because of this, one might be tempted to dismiss any interpretation of this phenomenon in terms of ``gravitational mass". However, since in the formalism of \cite{Nicolis:2017eqo} the effect is due to a coupling with gravity in the effective Lagrangian of the phonon, the same coupling must affect the field equation {\em for} gravity: the (tiny) effective gravitational mass of the phonon generates a (tiny) gravitational field. The source of this gravitational field travels with the phonon. This point was not appreciated in~\cite{Nicolis:2017eqo}. 

Thus, in a very physical sense, the phonon carries (negative) mass. Moreover, this is not due to the usual equivalence of mass and energy in relativity: the effect survives in the non-relativistic limit. And, finally, it is not a quantum effect, because the formalism of  \cite{Nicolis:2017eqo} applies unaltered to classical waves.

In this paper we confirm this result by computing explicitly the mass carried by a classical sound wave packet. As we will see, 
from the wave mechanics standpoint the fact that such a mass is nonzero is a non-linear effect, and that is why from a linearized analysis we usually infer that sound waves do not carry mass. We also generalize the result to sound waves in ordinary fluids and to longitudinal and transverse sound waves in solids. We find that, in the non-relativistic limit, the  mass carried by a sound wave traveling in these media is its energy $E$ times a factor that only depends on the medium's equation of state:
\begin{align}  
M = - \frac{d \log c_s}{d \log \rho_m}  \frac{E}{c_s^2} \; , \label{M}
\end{align}
where $\rho_m$ is the medium's mass density and $c_s$ the speed of the sound wave under consideration. (For solids, the correct formula is a minor generalization of this.) The derivatives in Eq.~\eqref{M} are standard adiabatic ones.  Notice that the same logarithmic derivative as in \eqref{M} appears in the general formula for a sound wave's radiation pressure \cite{Brillouin}.

A final qualification: for excitations propagating in a Poincar\'e-invariant vacuum, the invariant mass is a fundamental, completely unambiguous  quantity, whose value also directly determines the gap. Not so in a medium that breaks boost invariance: there, the dispersion relation $E(p)$ of an excitation is not constrained by the symmetries anymore, and, as a consequence, there is no ``invariant mass". 
What do we mean by `$M$' in \eqref{M} then? 
A non-relativistic medium---i.e. one with mass density (times $c^2$) much bigger than its energy density---has a well-defined and conserved total mass. 
In this case, the mass $M$ is simply the fraction of this total mass that travels with the excitation. It is perfectly consistent with the symmetries of the system to have {\em gapless} excitations ($E(p) \to 0$ for $p \to 0$) that nevertheless carry mass\footnote{Because of this, we find the use of ``massive" instead of ``gapped" in \cite{Watanabe:2013uya} somewhat inaccurate.}.

For technical convenience, we use gravity to probe how much mass our excitations carry, since in the non-relativistic limit gravity only couples to mass.

%
%

\vspace{0.5em}

\noindent\emph{Conventions}: We set $\hbar=1$ but keep powers of $c$ (speed of light) explicit. We adopt a signature $\eta_{\mu\nu}=\text{diag}(-,+,+,+)$. Greek indices run over all spacetime coordinates, $x^\mu = (ct, \vec x)$, while Latin indices over spatial coordinates only.

\vspace{0.5em}

\noindent\emph{\textbf{{General framework and strategy}}} --- 
As a general framework, we use the recently developed effective field theories (EFT) for the gapless excitations of generic media. These can be characterized in terms of spontaneous symmetry breaking. In particular, all media break at least part of the fundamental spacetime symmetries of Nature, which are spacetime translations, spatial rotations, and Lorentz boosts (the Poincar\'e group) \cite{Nicolis:2015sra}.
The Goldstone modes associated with the spontaneous breaking of these symmetries are gapless, and correspond to the collective excitations of the system (see e.g.~\cite{Lange:1965,Leutwyler:1996,Lange:1966,Nicolis:2012vf,Greiter:1989,Son:2002,Hoyos:2013eha,Moroz:2014ska,Dubovsky:2006,Endlich:2011,Geracie:2014iva,Ballesteros:2016kdx,Ballesteros:2014sxa,Gripaios:2014yha,Leutwyler:1996,Dubovsky:2006,Endlich:2011,Endlich:2013, Alberte:2018doe}). 

Here we compute the mass carried by a sound wave packet. For each of the media considered, we use the corresponding effective action for gapless excitations to derive the equations of motion by varying the action w.r.t.~the fields, and the energy-momentum tensor $T_{\mu\nu}$ by varying the action w.r.t.~to the spacetime metric. Gravity interacts with any system through the system's energy-momentum tensor, and, in particular, in the non-relativistic limit the mass of a wave packet will be the spatial integral of $\frac{1}{c^2}T_{00}$ evaluated on a solution of the equations of motion. 
Working perturbatively in the amplitude of our waves, we keep the linear terms and the leading non-linear ones, since the effect we are interested in vanishes at linear order.

Finally, we keep a relativistic notation and take the non-relativistic limit only when needed. It should  be noted that the same results can also be derived imposing Galilean invariance from the beginning. However, the relativistic analysis is not any more demanding to carry out, and it is more general. 
Once powers of $c$ are made explicit, the ratio of the speed of sound to $c$ encodes how relativistic the system is. We will see that our effect survives when this ratio tends to zero, so it is not associated with the usual relativistic equivalence of mass and energy.

\vspace{0.5em}

\noindent\emph{\textbf{Zero-temperature superfluids}} ---
The low-energy dynamics of a zero-temperature superfluid can be characterized by a single scalar field $\phi(x)$, with low-energy effective action~\cite{Greiter:1989, Son:2002}
\begin{equation} \label{eq:actionsuperfluid}
  S = \int dtd^3 x\, P(X) \; , \qquad X \equiv - c^2 \partial_\mu \phi \partial^\mu \phi \, ,
\end{equation}
expanded about the ground state $\langle \phi(x)  \rangle = \mu t$, where $\mu$ is the (relativistic) chemical potential. The function $P$ is the pressure of the superfluid.

The superfluid phonon field $\pi(x)$ corresponds to fluctuations of the scalar around its background, $\phi(x)=\mu(t+\pi(x))$. The e.o.m.~associated with the action~\eqref{eq:actionsuperfluid} is $\partial _\mu (P'(X) \partial^\mu \phi) =0$, which expanded to quadratic order  in small fluctuations reads
\begin{align}
  \label{eq:sfEOM}
  \begin{split}
  \ddot{\pi} - c_s^2 \nabla^2 \pi &=c^2\left(1-\frac{c_s^2}{c^2}\right)\dot\pi\nabla^2\pi \\
  &\;\;\;\; +\partial_t\left[c^2\left(1-\frac{c_s^2}{c ^2}\right)\big(\vec\nabla\pi\big)^2-g\dot{\pi}^2\right]\,.
  \end{split}
  \end{align}
The speed of sound $c_s$ and non-linear coupling $g$ are given by combinations of derivatives of $P(X)$, evaluated on the background field $\phi(x) = \mu t$. They can be rewritten in terms of the chemical potential as $c_s^2 = c^2 \frac{dP/d\mu}{\mu \, d^2 P/d \mu^2}$ and  $g = \frac{c^2}{2c_s^2} \left( 1 - \frac{c_s^2}{c^2} \right) - \frac{\mu}{c_s} \frac{dc_s}{d\mu}$.

We can solve this equation to non-linear order by writing $\pi = \pi_{(1)} + \pi_{(2)} + \dots$, where $\pi_{(1)}$ is a localized wave packet obeying the linear equation of motion,
\be
\ddot{\pi}_{(1)} - c_s^2 \nabla^2 \pi_{(1)} = 0 \; ,
\ee
and $\pi_{(2)}$ a quadratic correction, sourced by $\pi_{(1)}$ itself, 
 \begin{align}
\ddot{\pi}_{(2)} - c_s^2 \nabla^2 \pi_{(2)} & = \partial_t\left[ c^2\left(1-\frac{c_s^2}{c^2}\right)\big(\vec\nabla\pi_{(1)}\big)^2 -g\dot{\pi}_{(1)}^2\right. \notag \\
&\qquad\;\;\,+\left.\frac{c^2}{2c_s^2}\left(1-\frac{c_s^2}{c^2}\right)\dot{\pi}_{(1)}^2\right] \, . \label{pi2}
\end{align}

To find the energy-momentum tensor, we vary the effective action~\eqref{eq:actionsuperfluid} w.r.t.~the metric, which enters the spacetime volume element and the invariant $X = -c^2 g^{\mu\nu} \partial_\mu \phi \partial
_\nu \phi$
\footnote{As usual, non-minimal couplings with the spacetime curvature yield contributions to $T_{\mu\nu}$ in flat space that are suppressed by extra derivatives. In fact, they do not contribute at all to the global charges---spatial integrals of $T_{0\mu}$---that we are trying to compute.}. The result is
\begin{align}
  T_{\mu\nu} = P(X) \eta_{\mu\nu} + 2c^2 P'(X) \partial_\mu \phi \partial_\nu \phi \, .
\end{align}
Expanding $T_{00}$ to quadratic order in small fluctuations, we find
\begin{align}
  \label{eq:sfT00}
  \begin{split}
  T_{00} &= \frac{\mu n c^2}{c_s^2} \bigg[ \dot{\pi} + \frac{c^2}{2c_s^2} \left( 1 - \frac{2\mu c_s}{c^2} \frac{dc_s}{d\mu} \right) \dot{\pi}^2\\
  &\;\;\;\; - \frac{c^2}{2} \left( 1 - \frac{2c_s^2}{c^2} \right) \big( \vec{\nabla} \pi \big)^2 \bigg] \, ,
  \end{split}
\end{align}
where $n = dP/d\mu$ is the number density.

To find the mass transported by the sound wave, we integrate $T_{00}$ over a volume containing the wave packet and average it over a time interval $\tau$ much larger than the typical oscillation time, $\tau \gg 1/\omega$. 
The linear term in $T_{00}$ can receive contributions both from the linearized wave packet $\pi_{(1)}$ and from the non-linear correction $\pi_{(2)}$. The contribution from $\pi_{(1)}$ vanishes, since it is a total time derivative and we are integrating on a time interval over which the wave packet averages to zero. The contribution from $\pi_{(2)}$ vanishes as well: the non-linear source on the r.h.s.~of \eqref{pi2}, in general, contains both oscillatory terms as well as constant ones, but the time derivative gets rid of the latter. This means that $\pi_{(2)}$ is purely oscillatory as well, and hence its time derivative vanishes in our time average.

As far as the quadratic terms in $T_{00}$ are concerned, only the linearized wave packet $\pi_{(1)}$ can contribute to this order in perturbation theory. We can relate their volume integral and time average to the energy $E$ of our wave packet, that is, the value of the free Hamiltonian. The Hamiltonian density that characterizes the ground state of a zero-temperature superfluid as well as its excitations is $T^{00} - \mu J^0$, where $J^\mu = 2 P'(X) \partial^\mu \phi$ is the Noether current associated with particle-number conservation. We thus find
\be 
E =  \frac{\mu  nc^2}{c_s^2}  \int d^3 x \frac12 \left[ \dot{\pi}^2 + {c_s^2} \big( \vec{\nabla} \pi\big)^2 \right]\,.
\ee
Representing the volume integral and time average with $\langle \int \dots \rangle$, and using the virial theorem, we have
\begin{align} \label{eq:quadraticterm}
\la \sint \dot{\pi}^2 \ra = c_s^2 \la \sint\big( \vec{\nabla} \pi \big)^2 \ra = \frac{c_s^2}{\mu n c^2} \, E \, .
\end{align}
Putting everything together,
\begin{align}
\la \sint  T_{00} \ra = \Big(1 -\frac{\mu}{c_s}\frac{dc_s}{d\mu} \Big) E = \Big(1 - \frac{\rho + P}{c_s} \frac{d c_s}{d P} \Big) E \, ,
\end{align}
where in the last step we used the zero-temperature thermodynamic identities $\rho + P = \mu n$, $d P = n d \mu$. 

The above result is fully relativistic. In the non-relativistic limit, the mass carried by our wave packet is
\begin{equation}
  M = \frac{1}{c^2}\la \sint  T_{00} \ra \simeq - \frac{\rho_m}{c_s}\frac{dc_s}{dP} \, E = - \frac{\rho_m}{c_s}\frac{dc_s}{d\rho_m} \, \frac{E}{c_s^2} \; ,
\end{equation}
which is indeed Eq.~\eqref{M}, as obtained in~\cite{Nicolis:2017eqo} from the point-particle effective theory. 
It is simple to show that the same result can also be recovered in quantum field theory for single-phonon states.

\vspace{0.5em}

\noindent\emph{\textbf{{Solids and fluids}}} ---
As  mentioned in the Introduction, the bending of a sound waves's trajectory in the presence of gravity is equivalent to Snell's law.
In light of this, given the generality of Snell's law, we expect sound waves to interact with gravity in fluids and solids much in the same way as they do in superfluids. We will again analyze phonons in a solid from an effective field theory viewpoint. Moreover, since in the EFT language a fluid is simply a solid with an enhanced symmetry (see e.g.~\cite{Endlich:2013,Nicolis:2015sra}), the results of this section extend straightforwardly to fluids as well. The effective field theory of solids has been developed in~\cite{Dubovsky:2006,Endlich:2011,Endlich:2013}, and we will in large part follow their notation. 

The EFT for a three-dimensional solid can be built out of three scalar fields, $\phi^I$ ($I=1,2,3$), representing the co-moving coordinates of the solid's volume elements. In the ground state, they can be aligned with the physical coordinates, $\langle\phi^I(x)\rangle=b_0^{1/3} \, x^I$, where $b_0$ is a dimensionless free parameter representing the degree of compression/dilation of the solid. (The cubic root convention has been chosen for later convenience.)

The low energy effective action is a generic function of the combination $B^{IJ} \equiv \partial_\mu \phi^I \partial^\mu \phi^J$ that is invariant under the rotational symmetries of the solid, which act on the $I,J$ indices. For simplicity, we focus on solids that are isotropic at large distances\footnote{Generalizing our analysis to a solid with discrete rotational symmetry is straightforward, although quite tedious.}. In such a case, there are only three independent invariants that we can form out of $B^{IJ}$.
We find it convenient to work with
\begin{align}
  b \equiv \sqrt{\det B} \, , \quad Y \equiv \frac{\text{tr} B^2}{\left( \text{tr} B \right)^2} \, , \quad Z \equiv \frac{\text{tr} B^3}{\left( \text{tr} B \right)^3} \, .
\end{align}
The normalizations of $Y$ and $Z$ have been chosen so that $b$ is the only quantity that retains information about $b_0$---$Y$ and $Z$ are invariant under an overall rescaling of our solid.
The most general action for our solid is then
\begin{align} \label{eq:FbYZ}
  S = - w_0 \int dtd^3 x\, f(b,Y,Z) \, ,
\end{align}
where $w_0$ is a constant with the units of a mass density, which we pulled out for convenience. It is convenient to set $w_0$ to be $1/c^2$ times the value of the relativistic enthalpy density $\rho+P$ on the ground state. With such a normalization, on the background one has $\partial f /\partial b = c^2$~\cite{Endlich:2011}, which we implicity use below.

A perfect fluid can be viewed as an infinitely symmetric solid~\cite{Nicolis:2015sra,Endlich:2013}, whose action is invariant under volume preserving diffeomorphisms acting on the $\phi^I$ fields. 
Then, the only invariant in this case is $\det B$, and we can recover the fluid EFT from the solid one by simply setting to zero all derivatives of $f$ with respect to $Y$ or $Z$.

Sound waves once again correspond to the fluctuations of the fields about their backgrounds, $\phi^I(x) = {b_0}^{1/3} \left( x^I + \pi^I(x) \right)$.
Expanding the invariants for small fluctuations, one obtains the phonons' quadratic equations of motion
\begin{align} \label{eq:solEOM}
  \ddot{\pi}_i - c_L^2 \partial_i \vec{\nabla}\cdot\vec{\pi} - c_T^2 \big(\nabla^2 \pi_i - \partial_i \vec{\nabla}\cdot\vec{\pi}\big) = \partial_j A_{ij} + \partial_t B_i \; ,
\end{align}
where
\begin{align}
  \label{eq:Aij}
  \begin{split}
  A_{ij} &= g_1 \delta_{ij} \dot{\vec{\pi}}^2 + g_2 \dot{\pi}_i \dot{\pi}_j + 3g_3 \delta_{ij} \big( \vec{\nabla}\cdot\vec{\pi} \big)^2\\
  &\;\;\;\; + g_4 \delta_{ij} (\partial_k \pi_\ell)^2 + 2g_4 \partial_j \pi_i \vec{\nabla}\cdot\vec{\pi}\\
  &\;\;\;\; + g_5 \delta_{ij} \big( \vec{\nabla} \times \vec{\pi} \big)^2 + 2g_5 \partial_j \pi_i \vec{\nabla}\cdot\vec{\pi}\\
  &\;\;\;\; - 2g_5 \partial_i \pi_j \vec{\nabla}\cdot\vec{\pi} + g_6 \partial_i \pi_k \partial_j \pi_k + g_6 \partial_k \pi_i \partial_k \pi_j\\
  &\;\;\;\; + g_6 \partial_k \pi_i \partial_j \pi_k + 3g_7 \partial_i \pi_k \partial_k \pi_j \, ,
  \end{split}
\end{align}
and the vector $B_i$ is a quadratic combination of fields whose particular form is irrelevant for what follows\footnote{Note that we stopped differentiating between lower-case and upper-case spatial indices,  because the background configurations $\phi^I \propto x^I$ break spatial and internal rotations down to diagonal combinations, and so the equations of motion for fluctuations will only be invariant under such combinations.}.
The sound speeds and non-linear couplings are given by
\begin{subequations}\label{eq:defs}
\begin{align}  
  c_L^2 &= b_0 F'' + \frac{16}{27} \frac{f_Y + f_Z}{b_0} \, ,\\
  c_T^2 &= \frac{4}{9} \frac{f_Y + f_Z}{b_0} \, ,\\
  g_1 &= - \frac{1}{2} - \frac{c_L^2}{2c^2} + \frac{c_T^2}{c^2} \, ,\\
  g_2 &= 1 - \frac{c_T^2}{c^2} \, ,\\
  g_3 &= \frac{c_L^2-c_T^2}{3} + \frac{1}{3} b_0 c_L c_L' - \frac{10}{9} b_0 c_T c_T' + \frac{16}{243} \frac{f_Z}{b_0} \, ,\\
  g_4 &= - \frac{c_L^2}{2} + 2b_0 c_T c_T' - \frac{8}{27} \frac{f_Z}{b_0} \, ,\\
  g_5 &= \frac{c_L^2-c_T^2}{2}  - b_0 c_T c_T' + \frac{4}{27} \frac{f_Z}{b_0} \, ,\\
  g_6 &= c_T^2 + \frac{2}{9} \frac{f_Z}{b_0} \, ,\\
  g_7 &= \frac{2}{27} \frac{f_Z}{b_0} \, ,
\end{align}
\end{subequations}
where the prime, subscript $Y$, and subscript $Z$ denote, respectively, differentiation with respect to $b$, $Y$, and $Z$,  and all derivatives are evaluated on the background. Notice that, with the exception of $f'$, the derivatives of $f$ with respect to its arguments are of order $c_L^2$ or $c_T^2$, and not of order $c^2$ \cite{Endlich:2011}. This is what makes the $T_{00}$ below so simple in the non-relativistic limit.

Let us now find the gravitational mass carried by sound waves in a solid (or fluid).
It is convenient to take the non-relativistic limit right away. The only  $O(c^2)$ term in $T_{00}$ turns out to be
\begin{align} \label{eq:T00solid}
  T_{00} & \simeq c^2 w_0 b_0 \vec{\nabla}\cdot\vec{\pi}  \, .
\end{align}
In particular, with our parametrization of fluctuations, quadratic and higher-order terms contribute to the energy density of the solid, but not to its mass density.
The mass of a sound wave is therefore
\begin{align} \label{eq:mg}
  M = w_0 b_0 \la \sint \vec{\nabla}\cdot\vec{\pi} \, \ra \, .
\end{align}

The volume integral reduces to a boundary term and, for large enough volumes, one might be tempted to discard it. This is certainly allowed for a linearized solution, since we can take it to be as localized as we wish. However, in general we must be careful about non-linear corrections. Splitting as before the fluctuation field into $\vec \pi = \vec \pi_{(1)} + \vec \pi_{(2)} + \dots$, from \eqref{eq:solEOM} one sees that for a localized wave packet $\pi_{(1)}$,  the non-linear correction $\pi_{(2)}$ can scale as $1/r^2$ at large distances, thus potentially giving a finite contribution to \eqref{eq:mg}.

To compute such a contribution, we first take the divergence of \eqref{eq:solEOM}, obtaining a wave equation directly for $\Psi \equiv \vec \nabla \cdot \vec \pi_{(2)}$, with $\vec \pi_{(1)}$-dependent sources:
\be
(\partial_t ^2 -  c_L^2 \nabla^2) \Psi = \partial_i \partial_j A_{ij} + \partial_t (\vec \nabla \cdot \vec B) \,.
\ee
Time-averaging this equation, we get rid of the total time derivative terms:
\begin{align} \label{eq:poissonsolid}
  \nabla^2 \la \Psi \ra = - \frac{1}{c_L^2} \
  \la \partial_i \partial_j A_{ij} \ra \; ,
\end{align}
where $A_{ij}$ is understood to be a function of the linear solution $\vec\pi_{(1)}$. The resulting Poisson's equation is solved using Green's function methods, ultimately leading to the volume-integrated and time-averaged expression we are after\footnote{To get to this, we used the distributional identity 
\begin{align}
\partial_i \partial_j \frac1r = \frac1{r^3}(3 \hat r^i \hat r^j - \delta^{ij}) - \frac{4\pi}{3} \delta^3(\vec x) \delta^{ij} \; , \notag
\end{align}
and the fact that the first piece---which should be interpreted as a 3D principal part, that is, by excising an infinitesimal ball around $r=0$---integrates to zero in any spherical region centered at the origin.}:
\begin{align} \label{eq:Aii}
  \la \sint \vec{\nabla}\cdot\vec{\pi} \ra \simeq - \frac{1}{3c_L^2}  \la \sint A_{ii} \ra \; .
\end{align}

The right-hand side of \eqref{eq:Aii} can be evaluated straightforwardly from the definitions in~\cref{eq:defs,eq:Aij}. It is convenient to split $\vec{\pi}_{(1)}$ into longitudinal ($L$) and transverse ($T$) components, for which the energies read
\begin{subequations} \label{eq:pcLT}
\begin{align}
E_L&= w_0 b_0 \int d^3 x\, \frac{1}{2} \left[  \dot{\vec{\pi}}^2_L + {c_L^2} \big( \vec{\nabla}\cdot\vec{\pi}_L \big)^2 \right]  \, , \label{eq:pcL} \\
E_T&=w_0 b_0 \int d^3 x\, \frac{1}{2} \left[  \dot{\vec{\pi}}^2_T + {c_T^2}\big( \vec{\nabla}\times\vec{\pi}_T \big)^2 \right] \, , \label{eq:pcT}
\end{align}
\end{subequations}
and for which the obvious generalizations of the virial relation \eqref{eq:quadraticterm} hold.

Putting together~\cref{eq:mg,eq:pcLT,eq:Aii} and using the analogs of \eqref{eq:quadraticterm}, in the non-relativistic limit we find
\begin{align}
  M & \simeq - \frac{w_0 b_0^2}{c_L^2} \bigg[ c_L {c_L'} \la \sint \big(\vec{\nabla}\cdot\vec{\pi} \big)^2 \ra +{c_T c_T'} \la \sint \big( \vec{\nabla}\times\vec{\pi} \big)^2 \ra \bigg] \notag \\
  & = - \frac{b_0}{c_L^2} \bigg[ \frac{c_L'}{c_L} E_L + \frac{c_T'}{c_T} E_T \bigg] \; .
\end{align}

Longitudinal and transverse sound waves in general have different propagation speeds---in fact, one generally has $c_L^2 > \frac43 c_T^2$ \cite{landau1986theory}. So even if initially one has a wave packet made up of both components, over time these will split into two spatially separated wave packets, each propagating at its own speed. 
For each of these, one simply has
\begin{align}
  M = - \frac{d \log c_s}{d \log \rho_m} \frac{E}{c_L^2}\,,
\end{align}
where $c_s$ is the propagation speed for the type of sound wave in question ($L$ or $T$). We used the fact that in the non-relativistic limit the mass density is proportional to the level of compression, $\rho_m \propto b$~\cite{Dubovsky:2011sj}, and the derivative with respect to $\log \rho_m$ has to be taken in the absence of pure shear---that is, it corresponds to hydrostatic compression \cite{landau1986theory}.  (Using the standard relationships between the stress and strain tensors in an isotropic solid, one can also trade $d/d \log \rho_m$ for $K d/dP$, where $K$ is the bulk modulus.) This generalizes Eq.~\eqref{M} to the case of both longitudinal and transverse waves.

This result holds equally well for the sound waves in a perfect fluid, for which $f_Y=f_Z= c_T^2 = 0$, in which case it reduces directly to Eq.~\eqref{M}.
In fact, for a perfect fluid we also checked our result by directly taking the non-relativistic continuity and Euler equations for $\rho_m$ and a potential velocity field $\vec v = \vec \nabla \psi$, expanding them to quadratic order in the perturbations $\delta \rho_m$ and $\psi$ about a static and homogeneous background, and computing $M \equiv \int d^3 x \, \delta \rho_m$ for a perturbative solution of such equations of motion. We again obtain Eq.~\eqref{M}. We spare the reader the tedious details of that analysis.

\vspace{1em}

\noindent\emph{\textbf{{Conclusions}}} --- 
We showed that contrary to common belief, sound waves carry gravitational mass in a standard Newtonian sense: they are affected by gravity, but they  also \emph{source}  gravity. Our results show that this effect goes hand-in-hand with the non-linear interactions of sound, and that this is true for superfluids, fluids, and solids. In particular, for all these media, in the non-relativistic limit the mass transported by a sound wave is proportional to its energy times a coefficient that only depends on the medium's equation of state.

The  mass transported is in general quite small, of order $M \sim E/c_s^2$. For example, a very energetic phonon  in superfluid helium-4 with momentum $k\sim1$~keV (i.e.,  wavelength of order of the Bohr radius) carries a mass roughly of order $M\sim 1$~GeV, i.e. that of a single helium atom.
Nevertheless, it is possible to envision experimental setups where this effect could be detected. 

One possibility is to employ ultra-cold atomic or molecular gases. In these systems, not only might one be able to achieve very small sound speeds and enhance the effect, but one could also use suitable traps to simulate strong gravitational potentials~\cite{sebastian}. For example, for a cesium Bose-Einstein condensate, the typical sound speed is $c_s\simeq\sqrt{4\pi a_s n/m_\text{Cs}^2}$. The typical scattering length and density are $a_s\sim 100$~\AA~and $n\sim10^{13}$~cm$^{-3}$, leading to $c_s\sim10^4$~$\mu$m/s~\cite{sebastian2}. Assuming the maximum phonon momentum to be (conservatively) $k\sim1/a_s$, one gets a displaced mass of $M\sim 10^3$~GeV. For a condensate of radius $\sim 50$~$\mu$m, this is roughly $10^{-3}\div10^{-4}$ of the total mass~\cite{weber2003bose}. Although quite small, this could offer interesting perspectives for detection.

Another possibility might be to consider seismic phenomena. The wave generated by an earthquake of Richter magnitude $m=9$ carries an energy $E\sim10^{18}$~Joules which, for $c_s\sim 5$~km/s, corresponds to $M\sim10^{11}$~kg, and a change in gravitational acceleration $\delta g\sim10^{-4}$~nm/s$^2$. Atomic clocks and quantum gravimeters can currently detect tiny changes in the gravitational acceleration, up to fractions of nm/s$^2$~\cite{freier2016mobile,chou2010optical,Campbell90}. Given the rapid improvement of these techniques, one can imagine that in the not-too-distant future they will reach the sensitivity needed to detect the gravitational fields of seismic waves.

The effects we are considering may also be of relevance to neutron star dynamics, since gravity would affect phonon-mediated transport properties in the superfluid stellar interior~\cite{Migdal:1960,Ginzburg:1965}. The sound speed inside the core is expected to be relativistic~\cite{Zeldovich:1961,Haensel:2007}, but our results hold nonetheless, with minor generalizations.

Finally, since sound waves both source gravity and are affected by it, they can interact through gravity. In particular, for ordinary equations of state (higher sound speeds at higher pressures), their gravitational mass is negative. Still, since gravity is attractive for like charges, two sound wave packets running parallel to each other should start converging. Very, very slowly, of course.

\vspace{1em}

\begin{acknowledgments}
\noindent\textbf{\emph{Acknowledgments}} --- We are grateful to N.~Bigagli, G.~Chiriac\`o, S.~Garcia-Saenz, L.~Hui, A.~Lam, R.~Mcnally, R.~Penco, P.~Richards, I.~Z.~Rothstein, C.~Warner and S.~Will for illuminating discussions.  This work has been partially supported by the US Department of Energy grant de-sc0011941. The work done by A.E. is also supported by the Swiss National Science Foundation under contract 200020-169696 and through the National Center of Competence in Research SwissMAP.
\end{acknowledgments}

\bibliographystyle{apsrev4-1}
\bibliography{biblio}

\end{document}